\documentclass[12pt]{article}

\usepackage{epsfig}

\title{Experimental Consequences of
Time Variations of the Fundamental Constants}
\author{J.\,Rich \\ DAPNIA-SPP, CEA Saclay \\ 91191 Gif-sur-Yvette, 
France}

\begin{document} 

\maketitle                         

\begin{abstract}
We discuss the experimental consequences of 
hypothetical time variations of the fundamental constants.
We emphasize that from a purely phenomenological point of
view, only dimensionless fundamental constants have significance.
Two classes of experiments are identified that give results
that are essentially independent of the values of all constants.
Finally, we show that experiments that
are generally interpreted in terms of time variations of 
the dimensioned gravitional  constant $G$ are better interpreted
as giving limits on the variation of the dimensionless
constant $\alpha_G=Gm_p^2/\hbar c$.
\end{abstract}

\vspace*{10mm}


\setlength{\baselineskip}{4.0ex}

Evidence was recently reported by Webb et al. \cite{webb}
for a time variation of the fine-structure
constant $\alpha=e^2/4\pi\epsilon_0 \hbar c$.
The group compared the fine-structure splittings
of atomic absorption lines produced
by  high redshift intergalactic clouds  with the same splittings
produced by terrestrial atoms.
They found a slight difference in the splittings that
suggests that
$\alpha$ was lower in the past: 
$\Delta\alpha/\alpha=-0.72\pm 0.18 \times 10^{-5}$.
By ``past'' we mean of order
$10^{10}$ years ago when the light passed through the clouds.
This result is not confirmed by limits in variations of
$\alpha$ over shorter time scales \cite{Uzan} but this could
simply indicate a non-linear time variation.

The reported time variations of $\alpha$ 
have inspired a variety of theoretical
speculations, among them being theories
where the speed of light, $c$ varies with time, thus
inducing a variation of $\alpha\propto c^{-1}$ \cite{barrow,Moffat}.
This has generated a polemic because it is often
stated that only dimensionless constants like $\alpha$
have ``physical'' significance \cite{okun,Duff,Flambaum}.
It is perhaps not sufficiently emphasized that this is
due to the experimental nature of physics.
Generally speaking, experiments either 
count events or compare similarly dimensioned quantities.
For example, when a length is measured, one really measures
the ratio between the length in question and the length of
a standard ruler.  When an angle is measured, the ratio between
the lengths two sides of a triangle is converted to the angle through
trigonometry.
A velocity is measured by counting the number of ``ticks'' of
a clock as an object moves through a standard distance.
A reaction rate is given by counting the number
of events during the time that a standard clock gives
a standard number of ticks.
In all these experiments, only dimensionless numbers
are measured.
As such, experimental results  can only be sensitive to dimensionless
combinations of fundamental constants \cite{Cook}.

It is the purpose of this paper, to give a few
illustrations of this principle.
Following \cite{Turneare},
the strategy will be to estimate the complete dependence
of an experimental result on the fundamental constants by
including their effect on the structure of the experimental
apparatus.  Once this is done, we can see how the results
of an experiment will change with time if the fundamental 
constants change with time.
One of the surprising results will be that there are two
classes of experiments that yield results that are largely
independent of the values of the fundamental constants or
to their time variation.

It is most interesting to start with an experiment
that would be naively  expected to be sensitive to
time variations of $\hbar$.
One role of this constant is to relate ``particle''
properties like momentum to ``wave'' properties
like wavelength.
We therefore consider
the  double slit (Young) interference experiment
performed with non-relativistic electrons of momentum $p$.
The experiment is performed as in 
Fig. 1 with, electrons 
impinging upon a wall with 
two narrow slits  separated by a distance $d$.
Beyond the wall, one observes an interference pattern with
the angle between interference maxima, $\theta$, determined by $p$, $d$
and $\hbar$:
\begin{equation}
\theta \;=\; \frac{\lambda}{d} \;=\; \frac{2\pi \hbar }{pd}
\label{theta1}
\end{equation}
One is tempted to say that the angle depends on $\hbar$ through 
the numerator of  (\ref{theta1}) and that a time variation
of $\hbar$ would lead to a time variation of the measured
diffraction angle.
This is only part of the story since the distance $d$ 
is determined by the material of the wall whose structure depends
on the fundamental constants.
Interatomic spacings for solid materials are generally
of order the Bohr radius,
$a_{0}=4\pi\epsilon_{0}\hbar^{2}/m_{e}e^{2}$ where $m_{e}$ and $e$
are the mass and charge of the electron .
This is due to the fact that
$a_{0}$ is the only length that can be formed from
the three fundamental constants
that determine atomic structure: $\hbar$, $m_{e}$ and
$e^{2}/4\pi\epsilon_{0}$.
If the fundamental constants change, we can expect that
the ratio $d/a_0$ is relatively constant.  (For crystalline
materials, the ratio is basically the number of atomic sites along
the distance $d$).
We therefore  write the diffraction angle in the form
\begin{equation}
\theta \;=\;\frac{2\pi \hbar }{pa_0}\, \frac{1}{d/a_0}
\;=\;\frac{1}{p}\,
\frac{m_{e}e^{2}}{2\epsilon_{0}\hbar} 
\,\frac{1}{d/a_0}
\label{theta2}
\end{equation}

\begin{figure} 
\psfig{file=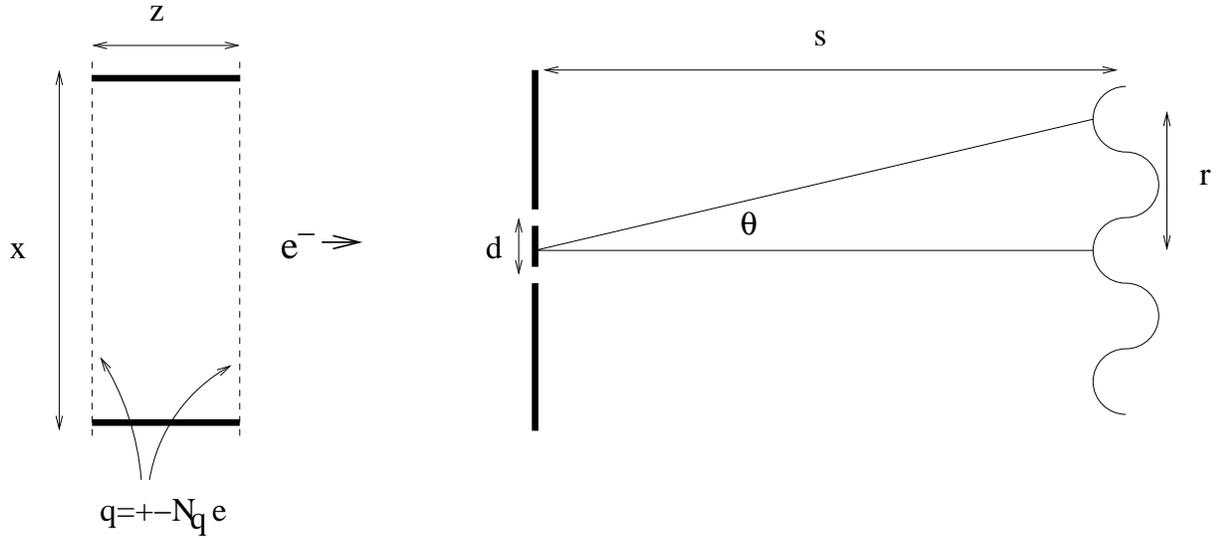,width=\textwidth}
\caption{
A double slit interference experiment.
Electrons are accelerated between two charged plates of area $x^2$ 
separated by a distance $z$.  
The electrons then impinge upon a wall having two narrow slits of
separation $d$.  The
angle $\theta$ between interference maxima is determined from the
ratio of the two distances $r$ and $s$.
As shown in the text,
if the distances $x$, $z$ and $d$ are determined by
solid objects, the diffraction angle is essentially independent
of the fundamental constants.
}
\end{figure}

It is amusing to note that whereas 
equation (\ref{theta1})
gives $\theta\propto\hbar$,
equation (\ref{theta2})
gives $\theta\propto\hbar^{-1}$.
However, we are not finished since nature does not usually provide
us with electrons of momentum $p$. Rather,  they must be  prepared
which can be done by passing electrons initially at rest
through a potential difference, $\phi$, so that 
$p^2/2m_{e}=e\phi$.
The potential difference can be maintained by placing a charge,
$\pm q$, on each of two square plates of area, $A=x^{2}$, separated
by a distance $z$.
This configuration gives $\phi=N_q ez/\epsilon_{0}x^{2}$
where $N_q=q/e$ is the number of fundamental charges per plate.
As with the interslit distance,
$x/a_0$  
and $z/a_{0}$ should we insensitive to changes in the fundamental
constants so we write the momentum in the form
$p^2/2m_{e}=N_{q}(e^2/\epsilon_0 a_0)(za_0)/x^2$.
Substituting this into  \ref{theta2}
we get the final result:
\begin{equation}
\theta = \sqrt{\frac{\pi}{2}}\,
         \frac{x/a_0}
              {d/a_0\sqrt{N_{q}z/a_0}} 
\label{theta3}
\end{equation}
To the extent that $x/a_0$, $d/a_0$ and $z/a_0$ do not
depend on the fundamental constants,
the interference angle does not depend on any
fundamental constant but only on the number and type
of the particles used to construct the experimental apparatus.      

This surprising result is, in fact, obvious
since an angle is a dimensionless
quantity and there is no
dimensionless combination of $\hbar$, $e^{2}/4\pi\epsilon_{0}$, 
and $m_{e}$ .
Measurements of angles using non-relativistic electrons and 
performed with
apparatuses whose size depend only on atomic structure
are therefore ``constant-free''.
Among such experiments are the Young experiment discussed
here and the Davisson-Thomson experiment on electron
diffraction by crystals. 
These quintessential quantum experiments give results
that are largely independent of the parameters of the theory 
(e.g. of $\hbar$) and 
depend only on the fact that the dynamics is governed
by quantum mechanics.
In the distant future when variations 
of the constants are routinely monitored
at the N.I.S.T., these experiments could serve are fiducial
experiments to check systematics.

Of course relativistic (fine-structure) corrections to interatomic
spacings would lead to a small dependence of the diffraction angles on
$\alpha$.
We also note that the time variation of a constant,
say of $e^2$, would introduce
a new constant $\tau_e=e/\dot{e}$ that gives the time
scale over which $e^2$ changes be a significant amount.
With this new constant we can now form the dimensionless
quantity $(e^2/\hbar)a_0/\tau_e$ which is just the fractional
change in $e$ during one Bohr revolution.
Depending on the dynamics that drives the variation
of the constants, the interference angle could 
conceivably depend on some power of this quantity.
On the other hand, 
if $\tau_e$ is of a cosmological scale, the parameter is tiny
and we might expect that
the diffraction angle is practically unaffected.

Our experiment to search for time variations of $\hbar$ can be
criticized because the size of the apparatus depends on $\hbar$.
We can avoid this criticism by
using the classical electron radius, 
$r_e=e^2/(4\pi\epsilon_0 m_e c^2)=\alpha^2 a_0$, 
to define the dimensions of the experimental
apparatus.  
We then
write (\ref{theta3}) as
\begin{equation}
\theta = \sqrt{\frac{\pi}{2}} \;\alpha^{-1} \;
         \frac{x/r_e}
              {d/r_e\sqrt{N_{q}z/r_e}} 
\label{theta4}
\end{equation}
The electron diffraction angle now depends on $\alpha$
if the distances are maintained as multiples of $r_e$.
Experimentally, this is difficult but
can be done by
filling a small cube of volume $l^3$ with a known number $N$
of electrons.
If the cube is uniformly irradiated by photons
of energy $E_{\gamma}\ll m_ec^2$,
the probability that a photon is scattered is 
$P=N\sigma_T/l^2$ where $\sigma_T=8\pi r_e^2/3$ is the
Thompson scattering cross section.  (We take $N$ small
enough so that $P\ll 1$.)
The probability $P$ can be measured so the length $l$ can be defined as
a multiple of $r_e$.
This length can then be compared with the lengths in
an experimental apparatus and an appropriate expansion-contraction
scheme can insure that all lengths remain a fixed 
multiple of $r_e$ even if the constants vary.

We now come  back to the original, more practical, experimental
arrangement using solid materials to define the apparatus.
The diffraction angle cannot depend on the constants because
there is no dimensionless combination of the relevant constants.
We can 
make $\theta$ depend on fundamental constants by accelerating  protons
instead of electrons.
In this case the angle is given by  (\ref{theta3}) multiplied
by $\sqrt{m_{p}/m_{e}}$ where $m_{p}$ is the proton mass.
The presence of two constants with dimensions of mass leads
to a new dimensionless quantity $m_{p}/m_{e}$ upon which the
angle can depend.

The interference angle can also depend on the fundamental constants
if we use photons, in which case we
we introduce the speed of light, c, into the problem.
For example, if we use photons from the $n=2$ to $n=1$ transition
of atomic hydrogen, we have a photon momentum of
$p_{\gamma}=(3/8)\alpha^{2}m_{e}c$
giving
\begin{equation}
\theta =
\frac{16\pi}{3} \alpha^{-1}
\frac{1}{d/a_0}
\label{thetag}
\end{equation}
A time variation of the fine-structure constant would then
yield a time dependence of the diffraction angle.

Having shown that a quintessential quantum experiment
gives results that are independent of $\hbar$,
we will now consider the quintessential relativistic effect,
the twin paradox.
Two identical clocks are needed, one in free fall, and the second
departing from the first with velocity $v$, stopping,  and then
returning to the first with the same velocity $v$.
The number of ticks on the two clocks, $N_1$ and $N_2$, 
counted during the time interval are related by
\begin{equation}
\frac{N_2}{N_1} \;=\; \sqrt{1-v^2/c^2} \;.
\end{equation}
We can now ask how this ratio will change if the fundamental
constants vary while maintaining the validity of the basic
formula.
The problem is to give a prescription for determining the 
velocity $v$.  If $v$ is determined by comparing the velocity
of the clock with $c$, we will clearly never detect any effect
since clocks of a given $v/c$ will always give the same $N_2/N_1$.
To detect an effect, we must define the velocity differently.
For instance, we can use clocks with the same velocity as
that of the electrons produced in Figure 1
\begin{equation}
\frac{v^2}{c^2} \;=\; 8\pi N_q \alpha^2 \, \frac{za_0}{x^2}
\end{equation}
This shows that 
if the experiment is built from  solid objects so that
$za_0/x^2$ is insensitive to changes in the fundamental constants,
then a time variation of $N_2/N_1$ should be interpreted as a 
time variation of $\alpha^2$.  On the other hand, if we design the
experiment so that $z$ and $x$ are fixed multiples of the
classical electron radius $r_e=\alpha^2 a_0$, then
the ratio $N_2/N_1$ is constant-free.
This is because there is no dimensionless
combination of the relevant constants, $e^2/4\pi\epsilon_0$, $m_e$
and $c$.
Just as non-relativistic quantum experiments give results
that are $\hbar$ independent, non-quantum relativistic
experiments can give results independent of $c$.

Since the twin paradox experiment cannot give unambiguous information
on the time variation of $c$,
we will now consider experiments
that directly measure $c$.
One might hope that a time variation of $c$ would lead
to a time variation
of the ratio between the flight time $x/c$ over a
distance $x$ and the period $T$ of a standard clock.
This amounts to counting the number of ticks $=(x/c)/T$ that the
clock makes during the time the photon travels the distance $x$.

If we define the distance $x$ by photon time-of flight as is
now done in SI units, we will obviously not be able to detect
a change in $c$.  We will then go back to the old procedure
of defining lengths in terms of physical rods.
We therefore write
\begin{equation}
\frac{x/c}{T} \;=\; \frac{a_0/c}{T} \; \frac{x}{a_0} 
\;=\; \frac{\hbar}{\alpha m_e c^2 T} \; \frac{x}{a_0}
\label{xcTbasic}
\end{equation}
and suppose that the ratio $x/a_0$ is insensitive to changes
in the constants.

The ratio $x/cT$ will now depend only on the clock that we choose
to use.  
Atomic clocks based on hyperfine splittings of
atomic lines have periods of order
\begin{equation}
T_{hf} \;\propto\; \frac{\hbar}{\alpha^4 g(m_e/m_p) m_e c^2}
\end{equation}
where g is the nuclear gyromagnetic ratio.
This gives
\begin{equation}
\frac{x/c}{T} \;=\; 
g \alpha^3 \,(m_e/m_p) \; \frac{x}{a_0}
\end{equation}
If $x/a_0$ is constant-free,
this attempt to detect a change in $c$ yields the time variation
of $g \alpha^3m_e/m_p$.
It is basically the technique of Turneaure and Stein \cite{Turneare}
who, by looking for the  desychronization 
of a superconducting cavity oscillator and
an atomic clock, set an upper limit of $4.1\times 10^{-12}yr^{-1}$
on the logarithmic derivative of $g \alpha^3 m_e/m_p$.

The use of a clock based on hyperfine splittings 
to measure variations of $c$ can be 
criticized because the clock period depends 
on a relativistic effect and therefore explicitly
on $c$.  It is possible to find quantum processes that
have periods that are $c$-independent.
The first type gives periods that are multiples of the
rotational period of an electron in a Bohr orbit:
\begin{equation}
T_e \;=\; \frac{a_0}{e^2/4\pi\epsilon_0 \hbar}
\;=\; \frac{\hbar}{\alpha^2 m_e c^2}
\label{Te}
\end{equation}
An example of such a clock is a mechanical vibrator.
The sound speed of a crystal consisting of
nuclei of mass $\sim Am_p$ is of order 
$(a_0/T_e)(m_e/Am_p)^{1/2}$ so a 
rod of length $L$ has a fundamental period 
$\propto (m_p/m_e)^{1/2}(L/a_0)T_e$.
The use of such a clock would give
\begin{equation}
\frac{x/c}{T} \;=\;
\alpha \;(m_e/m_p)^{1/2}\;\frac{x}{L}
\end{equation}
Since $x/L$ is constant-free for material objects,
a variation of the  measured speed of light would then be
interpreted as a variation of the quantity 
$\alpha(m_e/m_p)^{1/2}$.

A second class of $c$-independent periods can be found by
replacing $e^2/4\pi\epsilon_0$ with $Gm_p^2$ and $m_e$ by $m_p$:
\begin{equation}
T_G \;=\; \frac{\hbar^3}{G^2m_p^5}
\;=\; \frac{\hbar}{\alpha_G^2 m_p c^2}
\label{Tg}
\end{equation}
where $\alpha_G=Gm_p^2/\hbar c$ is the gravitational equivalent
of the fine-structure constant.
An example of a clock whose period is a multiple of $T_G$
uses a particle revolving near the surface of a totally
degenerate (white dwarf) star consisting of $N_e$ electrons
and $N_p$ nucleons.  
The radius of such a star is $R\sim\hbar^2N_e^{5/3}/(Gm_e m_p^2N_p^2)$ and
the particle has a revolution period of
\begin{equation}
T \;\sim\;   N_p^{-1} \left( \frac{m_p}{m_e}\right)^{3/2}
\left( \frac{N_e}{N_p}\right)^{5/2} \; T_G
\end{equation}
Substituting this into (\ref{xcTbasic}) gives
\begin{equation}
\frac{x/c}{T} \;\sim\; 
\frac{\alpha_G^2}{\alpha} \left(\frac{m_e}{m_p}\right)^{1/2} N_p 
\left(\frac{N_p}{N_e} \right)^{5/2}
(x/a_0) \;.
\end{equation}
A time variation of the measured speed of light using this clock
would be interpreted as a variation of 
$(m_e/m_p)^{1/2}\alpha_G^2/\alpha$.

We end with a comment on 
searches for time variations of the dimensional gravitational
constant $G$.
One way to do this is to search for anomalies in the
movement of objects in the Earth's gravitational field \cite{Uzan}.
The period $P$ of an object in a circular orbit of radius
$r$ is given by
\begin{equation}
P^2 \;=\; \frac{4\pi^2 r^3}{GM_{\odot}} \;=\; 
 \frac{4\pi^2 r^3}{GN_{\odot}m_p}
\label{periodorbit}
\end{equation}
where in the second form we approximate the solar mass by
$N_{\odot}m_p$ where $N_{\odot}$ is the number of nucleons
in the Sun.
If $G$ varies in time we can expect $P$ to vary in time.
Of course, there are two problems here.  First, what one would actually
observe is time variation of the ratio of $P$ and the period $T$
of a standard clock.  Second, a time variation of $G$ might
be expected to yield a time variation of the orbital radius.
If the radius changes, one must extrapolate the period back
to the original radius using (\ref{periodorbit}).
To do this, we need to measure 
$r$, which can be done with 
radar, $r=ct$, where $t$ is the time for a photon to
travel the distance $r$.  (We ignore small general relativistic
corrections.)  The time $t$ can then be fixed to be a given
number $N$ of periods of the standard clock, $r=cNT$.
The ratio of the orbital period (fixed $r$) to the
clock period  is now given by
\begin{equation}
\frac{P^2}{T^2} \;=\; 
\frac{4\pi^2 N^3}{N_{\odot}}\frac{c^3 T}{Gm_p} 
\;=\;
\frac{4\pi^2 N^3}{N_{\odot}}\frac{1}{\alpha_G \alpha^4 g(m_e/m_p)^2}
\;,\end{equation}
where in the second form we use $T=T_{hf}$ appropriate for
atomic clocks.
An anomalous variation of an orbital period can then 
be interpreted as a variation of $\alpha_G \alpha^4 g(m_e/m_p)^2$.
While not claiming to have analysed all experiments used
to limit the variations of $G$, this suggests that these
limits can all be interpreted as limits on the variation of
$\alpha_G$ and some combination of $\alpha$ and various mass
ratios and gyromagnetic ratios.

Since non-gravitational methods yield limits on
$\alpha$,\ $m_e/m_p$ and $g$ \cite{Turneare,Uzan}, 
that are considerably stricter than the limits on time
variations based on orbital anomalies,
an observed variation of an orbital period can be 
safely interpreted simply as a variation of $\alpha_G$.
Traditionally, limits on orbital anomalies have been
interpreted as limits on the time variation of $G$ with
the caveat ``assuming all other constants non-varying.''

In summary, we see that only dimensionless parameters have
phenomenological significance.  It should however be emphasized
that theories are generally  expressed in ways that
use dimensioned parameters.  If  nature is
such that the dimensionless parameters objectively vary with
time, it is conceivable that the underlying dynamics is
most simply expressed as a time variation of one or more
dimensioned parameters.

I thank Michel Cribier, Jean-Louis Basdevant and Marc Lachi\`eze-Rey
for interesting discussions.

\end{document}